# Relationship between Type of Risks and Income of the Rural Households in the Pattani Province of Thailand

Pha-isah Leekoi[1], Ahmad Zafarullah Abdul Jalil[1] & Mukaramah Harun[1]

[1] School of Economics, Finance and Banking, Universiti Utara Malaysia, Kedah Darul Aman, Malaysia

Correspondence: Pha-isah Leekoi, School of Economics, Finance and Banking, Universiti Utara Malaysia, 06010 UUM Sintok, Kedah Darul Aman, Malaysia. E-mail: faisah6666@yahoo.com

**Abstract**

This study examines the relationship between type of risks and income of the rural households in Pattani province, Thailand using the standard multiple regression analysis. A multi-stage sampling technique is employed to select 600 households of 12 districts in the rural Pattani province and a structured questionnaire is used for data collection. Evidences from descriptive analysis show that the type of risks faced by households in rural Pattani province are job loss, reduction of salary, household member died, household members who work have accident, marital problem and infection of crops/livestock. In addition, result from the regression analysis suggests that job loss, household member died and marital problem have significant negative effects on the households' income. The result suggests that job loss has adverse impact on households' income. The implication of this is that the living standard of household will continue to deteriorate as large proportion of them could either not find job or lost their jobs. Therefore, an important policy suggestion is that government should formulate a policy that considers the creation of employment especially for the poor households with low-income particularly in the rural area. Also, government should provide an appropriate social security benefits program on which the affected population can rely on in case of problem such as sickness/accident/death of the household members. Concerning the marital problem in the households, an important implication to the policy maker is to formulate a policy or design strategy development principles of holistic family.

## 1. Introduction

Thailand is an economically developing country situated in the Southeast Asian region. It is surrounded by Loas, Myanmar, Vietnam and Malaysia. The major economic sectors of Thailand such as agriculture, industry, manufacturing and services have added tremendously to its economic growth. The structure of Thailand's economy is originally based on agriculture. Like most of other developing countries, agriculture in the rural areas of Thailand remains as one of the main sources of income for the majority of the population. In the rural areas of developing countries, there is also widespread of income diversification. Therefore, it is possible for the rural households in one area to have substantial low and volatile incomes relative to the average income of the nation as a whole (Rungruxsirivorn, 2007). Households in the rural areas can lessen risk faced in many ways: they can select crops that have their yields or prices showing low correlations; plant crops on scattered plots of land that are exposed to various weather shocks, use different techniques of production, or choose a blend of farm and non-farm occupations (Alderman & Paxson, 1992). Poor rural households may concentrate only on one activity with low risk and low return in a case where there are no other options rather than diversifying income to manage risk.

Seasonality of profit and agricultural production impact the lives of farmers as well as the lives of other people in their communities because almost all the workers in the rural areas rely on agriculture. Apart from the fact that most of the rural households engage in agricultural production, they still take part in other economic activities such as self-employment in small industrial and commercial activity, and a paid work from fields such as agriculture, trade and other services. Other activities, such as trades or service, are also linked to the main income generating activities of majority of the rural households (Carlos Andres Alpizar, 2007; Leekoi, Abdul Jalil & Harun, 2014). However, some factors have been considered responsible for the fluctuation in agricultural income from one season to another. These include weather variations and other environmental factors such as limited and uncertain rainfall, floods or pest infestation. In addition, risk in term of natural, health, social, economic, political and environmental could be associated with the high fluctuation in the household income. For example, in Thailand, most of the risk faced by households stem from income fluctuation. As pointed out by the National Statistical Office, the household average monthly income has been on a steady increase after the financial crisis. The rising average income shows that the households' welfare is improving. However, evidence on actual households from the Ministry of Finance Household Debt (MOF) Survey gives a different picture (Rungruxsirivorn, 2007).

In the light of the fluctuation in the income of household and the various risk exposed to by the household, this study examines the relationship between type of risks and income of the rural households in Pattani province of Thailand. The rest of the paper is organized as follows: The section 2 gives an overview of relevant literature. In section 3, the methodology as well as the data used is discussed. Section 4 discusses the findings of the study. Finally, section 5 concludes the paper.

## 2. Literature Review

Risk can be defined as potentially dangerous event that when it occurs is likely to cause a loss in individual or household welfare. If households are defenceless when faced with a risk, it can easily be ruined and driven into poverty (Chaudhuri, Jalan, & Suryahadi, 2002; Dercon, 2002; Harrower & Hoddinott, 2004).

Household faces various risks which are mainly categorized as natural risks, health risks, economic risks, life-cycle risks, social risks, political risks, and environmental risks. The natural risk can be classified into floods, typhoons, earthquakes, droughts, volcanic eruptions, hurricanes, landslide while health risks involve accidents, injury, illness, disability, famines, and epidemic. The life-cycle risks include death, old-age, family break-up, maternity, and birth while the social risks are classified as terrorism, gangsters, crime, riots, social upheaval, and civil war. While economic risks involve loss of jobs, harvest destruction, resettlement, bankruptcy or business failure, currency crisis, banking and financial crisis, balance of payment shock, and the terms of trade shock, the political risks consist of manifestations, discrimination, and riots. Finally, environmental risks are made up of nuclear disaster, air and water pollution, land degradation, and deforestation (Holzmann & Jorgensen, 2000).

Certainly, the uncommon high variability in the income of rural households has been documented in the literature. For example, in a study using a 10-year panel data for one of the International Crops Research Institute for the Semi-Arid Tropics (ICRISAT) villages in India, Townsend (1994) found high fluctuations in yields per unit of land for the dominant crops. The study shows that the coefficient of variation was 1.01 for castor, 0.70 for paddy and 0.51 for a sorghum/millet/pea intercrop. In another study, Townsend (1995) uses the ICRISAT data for India and finds that variability in income is very high in this area. The study suggests that the effects of the diversification techniques and other income strategies employed have been limited and insufficient. Also in India, Bliss and Stern (1982) indicate that farmers in Palanpur will adversely be affected by the 20 percent decrease in yields if the onset of production is delayed by two weeks. In the literature, it has been argued that there are relatively large labor income fluctuations among casual labor in the agricultural sector (Lipton & Ravallion, 1995). For example, the coefficient of variation in farm household income in the south of India was estimated to be 137 while that of United States white men in the late twenties was only 39 (Rosenzweig & Binswanger, 1993).

Seyi Olalekan, Olapade-Ogunwole, and Raufu (2011) examine the types of shocks experienced by the rural households of the Ogo-Oluwa Local Government Area of Oyo state, Nigeria using a multi-stage random sampling technique and a probit regression model. The results reveal that most rural household heads experience more of ecological shocks which are common to agricultural production such as incidence of crop pests and livestock diseases, drought and degraded land. It is indicated that these shocks have significant impact on the household heads that possess poor educational status and per capita income. This is reflected in their capability and possibility to cope up with the incidence. Rampini and Viswanathan (2009) conducted a study on U.S. households and found that poor households and financially constrained households are less insured against risks (eg., health risks or natural disaster) compared to richer and less financially constrained households. Another study, De Mey et al. (2012) examine the concepts of operational, financial, total farm and household risk, using Belgian FADN data for the period 2005-2008. The study further uses a stochastic simulation model on two typical Belgian dairy farms and finds that price, production and financial farm risks are likely to have considerable negative effects on household incomes, which farmers may not be aware of.

## 3. Methodology

A multi-stage sampling technique was employed to select 600 households of 12 districts in the rural Pattani province and a structured questionnaire was used for the cross-sectional data collection. A descriptive statistic and standard multiple regression analysis were employed. Ordinary Least Squares (OLS) regression is a generalized linear modeling technique which is used to estimate the relationship between the true populations of two variables. The technique can be applied to single or multiple explanatory variables and also categorical explanatory variables that have been appropriately coded (Hutcheson, 2011). Given the assumption of the classical linear regression model, an OLS estimator is said to be a best linear unbiased estimator (BLUE) (Verbeek, 2004). Formal tests are conducted to examine whether the estimated results suffer from autocorrelation, heteroskedasticity and multicollinearity problem. In view of the fact that OLS estimator may suffer from the problem of autocorrelation and heteroskedasticity, a Newey-West estimator is further considered as an alternative for the regression. The estimator can be used to improve the OLS regression when the variables have heteroskedasticity and autocorrelation. Newey-West standard error known to be heteroskedasticity and autocorrelation consistent (HAC) has the advantage of being consistent in models that exhibit higher order autocorrelated errors. The estimator does not need the model of the error to be specified in a dynamic form which would have been required as an alternative to estimate the parameters more precisely (Newey & West (1987). The form of standard multiple regression equation used in this study is presented as such:

$$Y = \alpha + \beta_1\chi_1 + \beta_2\chi_2 + \beta_3\chi_3 + \beta_4\chi_4 + \beta_5\chi_5 + \beta_6\chi_6 + \varepsilon \qquad (1)$$

Where, Y= household's income, $\alpha$ = the constant of the equation, $\beta_{1-6}$ = the coefficients of the each predictor variables, $\varepsilon$ = error term, $\chi_1$ = job loss, $\chi_2$ = reduction of salary, $\chi_3$= household members died, $\chi_4$ = household members who work have accident, $\chi_5$ = marital problems (divorce, family dispute etc.), $\chi_6$ = infection of crops/livestock.

## 4. Result and Discussions

Table 1. Selected personal socio-economic variables of households' heads

| Variable | Frequency | Variable | Frequency |
|---|---|---|---|
| Gender | | Monthly households' income (Baht) | |
| Male | 416(69.3) | <5000 | 24(4.0) |
| Female | 184(30.7) | 5000-9999 | 61(6.2) |
| | | 10000-19999 | 268(44.7) |
| Age (years) | | 20000-29999 | 152(25.3) |
| 21-30 | 77(12.8) | 30000-39999 | 52(8.7) |
| 31-40 | 151(25.2) | 40000-49999 | 30(5.0) |
| 41-50 | 190(31.7) | >50000 | 13(2.2) |
| 51-60 | 122(20.3) | | |
| >60 | 60(10) | Households' assets | |
| | | Land | 271(30.2) |
| Marital status | | Shop | 87(9.7) |
| Married and living together | 459(76.5) | Rental house | 26(2.9) |
| Married but separated | 15(2.5) | Cars | 346(38.7) |
| Widowed | 112(18.7) | Livestock | 165(18.4) |
| Divorced | 14(2.3) | | |
| | | Types of risks | |
| Occupation | | Job loss | 37(21.6) |
| Farmers/agriculturists/fishery | 312(52.0) | Reduction of salary | 72(42.1) |
| Government officials | 86(14.3) | Household member died | 24(14.0) |
| Private sector employees | 9(1.5) | Household member who work have accident | 9(5.3) |
| Construction/labor/housemaid | 53(8.8) | Marital problem | 19(11.1) |
| Trade/own business | 140(23.3) | Infection of crops/livestock | 10(5.8) |

Source: Author's calculation from survey data, 2013, figures in parentheses are percentages.

Descriptive analysis was done to show the frequency distribution and percentage of some selected personal socio-economic variables of households' heads. Table 1 shows the results of the descriptive statistics. The results from the Table 1 show that most of the household head are male (about 69 percent) while the female head represents 31 percent. Most of the household heads in the sample whose ages fell within the range 41-50 years old represent about 32 percent and only 10 percent of the household head were aged above 60 years old. For the marital status of the household head, 459 respondents or 76.5 percent were married and living together while 2.3 percent of them were divorced. In term of employment types, almost half of the household head sampled reported to be farmers/agriculturists/fishery whereas only 1.5 percent of the total sample engaged in private sector employment. With regards to the income of the households, majority of the households sampled reported to have an income between 10,000-20,000 Baht per month (44.7 percent) while only 2.2 percent claimed to have earning more than 50,000 Baht per month.

In term of households' asset, most of the respondents have several forms of assets. Almost 40 percent of the respondents own a car while about 30 percent own a land. About 18 percent claim to possess livestock and 10 percent possess a shop. There were also a few of the respondents who have a rental house (about 3 percent). The results also indicate that the households encounter various risks. For example, reduction of salary is the risk that affected most (42 percent) of the households while only few (about 5 percent) household member who work has accident during the period of the study.

In order to determine the type of risks which affect a household's income, a standard multiple regression was conducted using the least squares method and the Newey-West standard error estimator. The results of the estimations are presented in Table 2.

Table 2. Regressions results for the impact of risks on the households' income

| Variable | Coefficient | OLS Std. | P >\|z\| | Newey-Wes | P >\|z\| |
|---|---|---|---|---|---|
| Constant | 10.722 | 0.323 | 0.000 | 0.316 | 0.000 |
| Job loss($\chi_1$) | -0.107 | 0.050 | 0.034** | 0.056 | 0.056*** |
| Reduction of salary($\chi_2$) | -0.043 | 0.152 | 0.776 | 0.116 | 0.710 |
| HHmmbr died($\chi_3$) | -0.327 | 0.052 | 0.000* | 0.062 | 0.000* |
| HHmmbr who work have acdt($\chi_4$) | 0.022 | 0.052 | 0.667 | 0.053 | 0.671 |
| Marital problems($\chi_5$) | -0.135 | 0.054 | 0.013** | 0.058 | 0.021** |
| Infection of crops/livestock($\chi_6$) | -0.111 | 0.126 | 0.378 | 0.105 | 0.290 |
| Number of observations | | 600 | | 600 | |
| F (6, 593) | | 10.02 | | 6.27 | |
| Prob> F | | 0.0000 | | 0.0000 | |
| R-squared | | 0.0829 | | | |
| Multicollinearity test: | | Mean VIF = 1.04 | | | |
| White's test for heteroskedasticity: | | Prob> $chi^2$ = 0.0623 | | | |
| Breusch-Godfrey LM test for autocorrelation: | | Prob> $chi^2$ = 0.0000 | | | |

Source: Author's calculation from survey data, 2013

Note: * denotes significance at 1 percent level, ** denotes significance at 5 percent level and, *** denotes significance at 10 percent level

Attempt was made by running the least squares regression first and this was followed by conducting the necessary tests to ensure better and reliable parameter estimates. From the least squares regression results in Table 2, the mean value of Variance Inflation Factor (VIF) for all variables is estimated to be 1.04. Since the value is less than 10, it implies that there is no multicollinearity problem in the model. The White test for heteroskedasticity yields $Chi.^2$ (21) = 31.73; Prob>$chi^2$ = 0.0623 indicating that there is no problem of heteroskedasticity at 5 percent level of significance. However, the Breusch-Godfrey test for autocorrelation shows that the estimation suffers from autocorrelation problem (Prob> $chi^2$ = 0.0000) at 5 percent level of significance. Based on the Gauss-Markov assumption of linear regression model, it is to recognize that the estimates from OLS are unbiased and consistent; it is just the standard errors that could be wrong (Verbeek, 2004). In order to correct for the problems associated with the least squares estimator, we attempt to employ the alternative approach namely, Heteroskedasticity and Autocorrelation Consistent (HAC) standard errors or Newey-West standard errors. The results of HAC or Newey-West estimator which corrects for the standard errors and produces robust estimates are reported in Table 2. The results imply that the heteroskedasticity and autocorrelation problems in the model are corrected.

Considering P >|z| values for the variables included in the model for estimation under the Newey-West estimators, the results show that job loss ($\chi_1$), marital problems ($\chi_5$), and household member died ($\chi_3$) are statistically significant at 10 percent, 5 percent and 1 percent level respectively. This simply indicates that an increase in the level of any of these explanatory variables definitely lead to a decrease in the level of households' income. For example, the coefficient of job loss ($\chi_1$) is -0.107; this implies that for every unit change in job loss, the households' income decreases by 10.7 percent. The household head or household member is meant to support the family with income earned. Therefore, an increase in job loss for this household will constitute a problem to the family due to its adverse implication on the income earned. The coefficient of marital problem is ($\chi_5$) is estimated to be -0.135. This indicates that for every unit change in marital problem, the households' income decrease by 13.5 percent. This is because when marital problems such as family dispute occur in the family, this leads to divorce and thus, no bread winner or income earner to support the households again. Furthermore, the coefficient of households member died ($\chi_3$) is estimated to be -0.327 which indicate that for every unit change in death of households member, the households' income decreased by 32.7 percent. This means that the incomes of the households dropped due to the death of the household members who generated income for the family. These findings are consistent with the previous result reported by Tesliuc and Lindert (2004). As for the reduction of salary ($\chi_2$), household members who work have accident ($\chi_4$) and infection of crops/livestock ($\chi_6$), they are found to be statistically insignificant and thus, have no effect on the households' income.

## 5. Conclusion

This study investigated the impact of risks on the households' income. The results of the descriptive analysis suggest that the risks faced by households in rural Pattani province are job loss, reduction of salary, household member died, household member who work have accident, marital problem and infection of crops/livestock. Among these six types of risk experienced by the households, only three (job loss, household member died and marital problem) have significantly impacts on the households' income.

The result suggests that job loss has adverse impact on households' income. The implication of this is that the living standard of household will continue to deteriorate as large proportion of them could either not find job or lost their jobs. Therefore, an important policy suggestion is that government should formulate a policy that considers the creation of employment especially for the poor households with low-income particularly in the rural area. Also, government should provide an appropriate social security benefits program on which the affected population can rely on in case of problem such as sickness/accident/death of the household members. Concerning the marital problem in the households, an important implication to the policy maker is to formulate a policy or design strategy development principles of holistic family. In addition, such strategy could entail the principles of ensuring social protection and creation of mechanisms to support and strengthen the family. Government and policy makers should endeavor to understand the general condition of households across the rural area in order to come up with an appropriate policy that will benefit the generality of the poor population.

This study recommends future studies to include other likely types of risk that could be peculiar to their rural area and which could probably have impact on the households' income to provide better results. Secondly, it would also be interesting if future study could expand the scope (the impact of risks on households' income) of the present study to other developing economies and different rural area, and compare their findings with that of Thailand. Such is likely to provide other good examples of several risks effect on the households' income.

**Copyrights**